\documentclass[a4paper,twocolumn,accepted=2020-03-12]{quantumarticle}
\pdfoutput=1

\usepackage[utf8]{inputenc}
\usepackage[T1]{fontenc}
\usepackage{bbm, amsmath, amssymb, amsthm, bm,textcomp, nicefrac,geometry,ragged2e}
\usepackage{dsfont}
\usepackage{graphicx,epstopdf,color,verbatim,enumitem,ulem,etoolbox}
\usepackage[dvipsnames]{xcolor}
\usepackage[bbgreekl]{mathbbol}
\usepackage[unicode=true,pdfusetitle, bookmarks=false,bookmarksnumbered=false,
bookmarksopen=false, breaklinks=false,pdfborder={0 0 0},backref=false,
colorlinks=true, linkcolor=myrefcolor,citecolor=myurlcolor,urlcolor=myurlcolor]{hyperref}
\usepackage{pbox,array}
\usepackage{algcompatible}
\usepackage{algpseudocode}
\usepackage{algorithm}
\usepackage{diagbox}
\usepackage[numbers,sort&compress]{natbib}
\definecolor{myrefcolor}{rgb}{0.8,0,0}
\definecolor{myurlcolor}{rgb}{0,0,0.7}

\makeatletter
\appto\appendix{%
  \@ifstar{\def\thesection{\unskip}\def\theequation@prefix{A.}}%
          {\def\thesection{\Alph {section}}}%
}
\makeatother

\DeclareSymbolFontAlphabet{\mathbb}{AMSb}
\DeclareSymbolFontAlphabet{\mathbbl}{bbold}

\newcommand{\ket}[1]{\left|#1\right\rangle}

\newcommand{\bra}[1]{\left\langle #1\right|}

\newcommand\ketbra[1]{|#1\rangle\langle#1|}
\newcommand{\proj}[1]{\ket{#1}\!\bra{#1}}
\newcommand\cH{{\mathcal H}}
\newcommand\cL{{\mathcal L}}

\newcommand\s{\text{s}}

\newcommand\id{\mathbbl{1}}

\newcommand*\colvec[1]{\begin{pmatrix}#1\end{pmatrix}}

\newcommand{\be}{\begin{equation}}
\newcommand{\ee}{\end{equation}}
\newcommand{\bea}{\begin{eqnarray}}
\newcommand{\eea}{\end{eqnarray}}

\newcommand{\fig}[1]{Fig.~\ref{#1}}
\newcommand{\alg}[1]{Alg.~\ref{#1}}
\newcommand{\eq}[1]{Eq.~(\ref{#1})}

\newcommand{\App}[1]{App.~\ref{#1}}

\def\tr{\mathrm{tr}}

\begin{document}

\title{What is the minimum CHSH score certifying that a state resembles the singlet?}
\author{Xavier Valcarce$^{1,*},$ Pavel Sekatski$^{1,*},$ Davide Orsucci$^{1},$ \\ Enky Oudot$^{1,2},$ Jean-Daniel Bancal$^{1,2}$ and Nicolas Sangouard$^{1}$}
\affiliation{$^1$ Departement Physik, Universit\"at Basel, Klingelbergstra{\ss}e 82, 4056 Basel, Schweiz \\
$^2$ D\'epartement de Physique Appliqu\'ee, Universit\'e de Gen\`eve, 1211 Gen\`eve, Suisse\\
$^*$ These authors contributed equally}
\begin{abstract}
    A quantum state can be characterized from the violation of a Bell inequality. The well-known CHSH inequality for example can be used to quantify the fidelity (up to local isometries) of the measured state with respect to the singlet state. In this work, we look for the minimum CHSH violation leading to a non-trivial fidelity. In particular, we provide a new analytical approach to explore this problem in a device-independent framework, where the fidelity bound holds without assumption about the internal working of devices used in the CHSH test. We give an example which pushes the minimum CHSH threshold from $\approx2.0014$ to $\approx2.05,$ far from the local bound. This is in sharp contrast with the device-dependent (two-qubit) case, where entanglement is one-to-one related to a non-trivial singlet fidelity. We discuss this result in a broad context including device-dependent/independent state characterizations with various classical resources.  
\end{abstract}

\maketitle 


\section{Introduction}
Entanglement and measurement incompatibility are two central properties of quantum physics. In 1964, John Bell pointed out that these features imply out-of-the-ordinary observable phenomena~\cite{Bell1964}. In particular, performing local incompatible measurements on subsystems of a state can lead to outcomes having stronger-than-classical correlations, also referred to as non-local correlations or non-locality. Interestingly, we can reconstruct the state and measurements (up to local isometries) generating these non-local correlations. This procedure is known as \textit{self-testing}~\cite{Mayers2004,Supic2019} and has the nice feature of being device-independent, i.e. the state is reconstructed without assumption on the internal working of devices used in the Bell test.

The simplest and most common Bell inequality is the Clauser-Horne-Shimony-Holt (CHSH) inequality~\cite{Clauser1969}. It applies to the scenario where two parties -- Alice and Bob -- share a state and perform each one out of two dichotomic (with two-outcomes) measurements. The experiment is repeated many times to assess the CHSH score - a single scalar computed from the statistics of the outcomes. Whenever the CHSH score is above 2, the underlying correlations between Alice and Bob's results cannot be described by a local causal theory. Furthermore, if the CHSH score attains the maximal quantum value of $2\sqrt{2},$ one can certify that Alice and Bob share a maximally-entangled two-qubit state, i.e. a singlet up to local unitaries~\cite{Popescu1992,Braunstein1992,Bardyn2009}. Hence CHSH \textit{self-tests} the singlet.

What can we say about the state structure when the CHSH score is below $2\sqrt{2}$? This is an important question in practice where imperfections unavoidably lead to non-maximal CHSH scores. Kaniewski showed that for all CHSH values above $\frac{2(8+7\sqrt{2})}{17} \approx 2.11,$ we can certify that the measured state has a non-trivial fidelity with respect to the singlet~\cite{Jed2016}. More precisely, if the CHSH score exceeds this value, there exists local extraction maps which output a
state, when applied to the actual state, with a fidelity larger than 1/2 with the singlet. Kaniewski also showed that the self-testing threshold for CHSH does not coincide with the local bound of 2. That is, there exist states violating the CHSH inequality for which there is no local map extracting a nontrivial singlet fraction~\cite{Jed2016}. In term of entanglement, this diverge from the two-qubit case where a trivial singlet fidelity implies separability.
Recently, an example of a state with a trivial singlet extractability has been found for a CHSH score of $\approx 2.0014$~\cite{Coopmans2019}. There is thus a \textit{threshold CHSH value} slightly higher than the local bound $2$ and smaller than $\approx 2.11$ below which it is not possible to do singlet self-testing. 

As of today, only few experiments are able to realize a proper Bell test, with no measurement cross-talk and no post-selection~\cite{Pironio2010,Christensen2013,Hensen2015,Giustina2015,Shalm2015,Rosenfeld2017,Liu2018}. The feasibility and requirements of self-testing with such setups strongly depends on the CHSH self-testing threshold. CHSH scores in photonic experiments using a source of polarization entangled pairs based on spontaneous parametric down conversion, for instance, are intrinsically limited by photon statistics and losses. As a result, a CHSH score of 2.11 can only be achieved with an overall detection efficiency above $\approx$90\%~\cite{CapraraVivoli2015}, way beyond what has been reported so far~\cite{Christensen2013,Giustina2015,Shalm2015,Liu2018}. Experiments with individual atomic systems can potentially deliver high CHSH scores~\cite{Pironio2010,Hensen2015,Rosenfeld2017} but they are currently limited by finite statistics. As far as we know, only two experiments so far~\cite{Pironio2010,Rosenfeld2017} can guarantee a mean CHSH score larger than 2.11 with a confidence of 99\%~\cite{Bancal2018}. A high CHSH self-testing threshold would thus pose a serious experimental challenge. On the other hand, if the threshold score for self-testing is close to the local bound, there must exist better extraction maps than the ones proposed in Ref.~\cite{Jed2016}. Such  maps could be used to significantly improve the robustness of a large class of self-tests. Discovering the threshold value for self-testing with the CHSH inequality, the most famous and commonly used Bell inequality, is thus both a natural and relevant question.

In this manuscript, we construct a two-qudit state (with a local dimension $d=6$) that has a trivial singlet extractability but nevertheless attains a CHSH value larger than $2.05.$ This is shown in Section \ref{CHSH_Selftesting_threshold} right after the preliminary Section \ref{Preliminaries}. Section \ref{Discussion} discusses this threshold in the broader context of device-dependent and device-independent state characterizations with various classical resources.  

\section{Preliminaries}
\label{Preliminaries}
\subsection{Self-testing}

We consider a scenario where Alice and Bob have access to a source of quantum states that produces copies of an unknown state $\rho_{AB}\in \cL(\cH_A\otimes \cH_B).$ They each have a measurement box with several setting choices delivering a measurement result in the form of a classical output. They ignore how the states are produced as well as the Hilbert space dimension. Similarly, they don't have a physical description of the measurement devices and don't know how they have been calibrated. For Alice and Bob, the source and measurement devices are thus black boxes. Nevertheless, they want to certify that $\rho_{AB}$ contains the target state $\Phi_{A'B'}^{+}$ from the sole knowledge of the measurement statistics.
Obviously, such a device-independent characterization can only be done up to local basis choices. Furthermore, Alice's and Bob's systems might not contain subsystems whose Hilbert space dimension matches that of the reference state. But Alice and Bob can identify the target state by applying local isometries $f_A$ anf $f_B$ on the state $\rho_{AB}$ distributed by the source \cite{Mayers2004}. An isometry is an embedding of a system into an Hilbert space of larger dimension, followed by a unitary transformation on the whole space. Concretely, the extractability of the reference state $\Phi_{A'B'}^+$ from the actual state can be defined as
\begin{align}
    \label{eq:extract1}
    \Xi[\rho_{AB} &\rightarrow \Phi_{A'B'}^{+}] = \\
    &\sup_{f_A, f_B,\varrho_\text{junk}} F((f_A \otimes f_B)[\rho_{AB}],\Phi_{A'B'}^{+}\otimes \varrho_\text{junk}),
    \nonumber
\end{align}
where $F(\rho_0,\rho_1)= \left(\tr[\sqrt{\rho_0^{1/2} \rho_1 \rho_0^{1/2}}]\right)^2$ is the square of the Uhlmann fidelity and $\varrho_\text{junk}$ is an irrelevant state of the auxiliary systems. One can think of isometries as the most general passive transformations. It is a mere basis choice in the big Hilbert space in which the system is embedded. Since the extractability is defined up to passive transformations, it tells one how much of the reference state \textit{is contained} in $\rho_{AB}$. 

Alternatively, the extractability can be defined after the auxiliary systems have been traced out. This comes from the fact that every quantum channel can be realized as a unitary acting on an extended space followed by a partial trace \footnote{In fact, any CPTP map $\Lambda: \cL(\cH)\to \cL(\cH_*)$ can be realized as a global unitary acting on $\cH\otimes \cH_*$, followed by discarding the $\cH$ subsystem.}. In this context 
Alice and Bob identify the reference state by applying local operations $\Lambda_A$ and $\Lambda_B$ on $\rho_{AB}$. It was shown in Ref. \cite[Proposition~2]{Sekatski2018} that the following definition of the extractability 
\begin{equation}
    \label{eq:extract}
    \Xi[\rho_{AB} \rightarrow \Phi_{A'B'}^{+}] = \sup_{\Lambda_A, \Lambda_B} F((\Lambda_A \otimes \Lambda_B)[\rho_{AB}],\Phi_{A',B'}^{+}),
\end{equation}
is equivalent to one given in Eq.~\eqref{eq:extract1}. Here $\Lambda_A : \cL(\cH_A) \rightarrow \cL(\mathds{C}^2)$ (resp. $\Lambda_B : \cL(\cH_B) \rightarrow \cL(\mathds{C}^2)$) is the extraction map -- a completely positive trace preserving (CPTP) map -- used by Alice (resp. Bob)\footnote{The reference Hilbert spaces can be different from $\mathds{C}^2$ in general.}. We note that  $F(\rho_0,\rho_1)=\tr[\rho_0 \rho_1]$ whenever at least one of the two states $\rho_0$ or $\rho_1$ is pure. In the following we will simply denote the extractability of $\rho_{AB}$ by $\Xi[\rho_{AB}]$, as there will be no ambiguity on the reference state
\be
\Phi^+_{A'B'}= \ketbra{\Phi^+}.
\ee
Notice that $\Xi[\rho_{AB}]$ is lower bounded by $1/2$. This is, for any state, the two parties can always 
apply a quantum channel that discards the received state and prepares e.g. $\ket{0}$. The resulting fidelity is then
\begin{equation}
    \Xi[\rho_{AB}] \geq F(\ketbra{0}\otimes\ketbra{0},\Phi^+_{A'B'})=\frac{1}{2}.
\end{equation}

\subsection{Self-testing from the CHSH score}
While complicated measurement devices can be envisioned, we here consider the simplest case with two setting choices -- $A_x$ for Alice, $B_y$ for Bob, with $\{x,y\}\in\{0,1\}$ -- and with two outcomes $\pm 1$. We define the operator
\be
W_{AB} = A_0\otimes (B_0 + B_1) + A_1\otimes(B_0-B_1) 
\ee
whose expected value gives the CHSH score
\be
S= \tr{\left(W_{AB}\rho_{AB}\right)}.
\ee
With a few steps, we can show that the singlet can be self-tested with the knowledge of $S$ only. Using Jordan's lemma~\cite{Supic2019}, we first choose a basis such that the observables $A_x$ and $B_y$ are simultaneously block diagonal with blocks of size two that are given by 
\be\label{eq: jordan}
\begin{split}
A_0(\alpha) &= Z, \qquad A_1(\alpha) =\cos(\alpha) Z + \sin(\alpha) X,\\
B_0(\beta) &= H_+, \qquad B_1(\beta) = \cos(\beta) H_+ + \sin(\beta) H_-.
\end{split}
\ee
$\alpha,\beta$ are angles belonging to $[0,\pi]$. $X,Y,Z$ denote the three Pauli matrices, $H_+=\frac{1}{\sqrt{2}}(Z+X)$ and $H_-=\frac{1}{\sqrt{2}}(Z-X).$ Trivially, the CHSH operator $W_{AB}$ inherits the block-diagonal structure, and thus can be expressed as a direct integral of two-qubit operators $W_{AB} = \int_{\alpha,\beta}^\oplus d\alpha d\beta \-\ W_{AB}^{\alpha,\beta}$. The quantum state $\rho_{AB}$ on which the measurements are performed does not a priori need to have the block-diagonal structure of the measurement operators. However, this can be assumed without loss of generality. Indeed, as a first step of their extraction map, Alice and Bob can always locally perform a projection onto their respective orthogonal blocks
\be\label{eq: block proj}
    \rho_{AB} \to \bar\rho_{AB} = 
    \int_{\alpha,\beta}^\oplus d\alpha d\beta \-\ p(\alpha,\beta) \rho_{AB}^{\alpha,\beta},
\ee
with $p(\alpha,\beta)$ the probability for a successful projection onto the blocks parametrized by $\alpha$ and $\beta.$ The extractability of $\rho_{AB}$ is thus larger or equal to that of $\bar \rho_{AB}.$ At the same time, the two states $\rho_{AB}$ and $\bar \rho_{AB}$ give the same CHSH score, as follows from the block-diagonal structure of $W_{AB}$. Hence, the state distributed by the source can be assumed to abide to the block diagonal structure of the observable $A_x$ and $B_y$, and be a mixture of two-qubit states $\rho_{AB}^{\alpha, \beta}$ across different blocks, that is
\be\label{eq: rho gen}
\rho_{AB} = 
\int_{\alpha,\beta}^\oplus d\alpha d\beta \-\ p(\alpha,\beta) \, \rho_{AB}^{\alpha,\beta}.
\ee
Then, the extraction maps can also be constructed following the block structure
\be
\Lambda_A^\alpha: \cL(\mathds{C}^2)\to \cL(\mathds{C}^2) \qquad \Lambda_B^\beta: \cL(\mathds{C}^2)\to \cL(\mathds{C}^2),
\ee
where the angles $(\alpha,\beta)$ specify the blocks. We note that whenever Alice and Bob receive the component $\rho_{AB}^{\alpha,\beta}$, Alice's map $\Lambda_A^\alpha$ can not depend on the value of the parameter of Bob's map $\beta,$ and similarly for Bob's map. Indeed, the values of angles are only available locally and contrary to e.g. entanglement theory, classical communication is not a passive transformation and thus cannot be used in self-testing. Hence, the singlet extractability for the state $\rho_{AB}$ in Eq.~\eqref{eq: rho gen} reads
\begin{align}
\Xi[\rho_{AB}] &=\\ \sup_{\{\Lambda_A^\alpha, \Lambda_B^\beta\} }&\int_{\alpha,\beta}\!\!\! d\alpha d\beta \-\ p(\alpha,\beta) \, F(\Lambda_A^\alpha\otimes\Lambda_B^\beta[\rho_{AB}^{\alpha,\beta}] ,\Phi^+_{A'B'})
\nonumber \end{align}
where the maximization is taken over all qubit maps $\Lambda^\alpha_A$, $\Lambda^\beta_B$ for which $p(\alpha,\beta)\neq 0$. Fixing the dependence of the maps on the parameters $\alpha$ and $\beta,$ we can lower bound the fidelity between the reference state $\Phi^+_{A'B'}$ and all two qubit states $\tau$ whose Bell score exceeds a certain value $S'$ by solving the following optimization 
\begin{eqnarray}
O_{\min}(S')=\min_{\alpha, \beta, \tau} \-\ && F(\Lambda_A^\alpha\otimes\Lambda_B^\beta[\tau] ,\Phi^+_{A'B'})\\
\nonumber
\text{s.t.} \-\ &&  \tr{\left(W_{AB}^{\alpha,\beta} \tau \right)} \geq S',\\
\nonumber
&& \tau \geq 0,\\
\nonumber
&& \tr{(\tau)} =1,\\
\nonumber
&& \tau^\dag = \tau.
\end{eqnarray}
The three last conditions ensure that $\tau$ is a physical state. For every $S',$ we run a semi-definite optimization for a clever choice of maps~\cite{Jed2016} over all possible states $\tau$ for each set of angles and then minimize the overlap over the angles. Finally, we take the convex roof of $O_{\min}(S)$ which gives a lower bound on the extractability $\Xi[\rho_{AB}]$. The result shows that a non-trivial extractibility can be obtained as long as $S \approx 2.11$~\cite{Jed2016,Sekatski2018}.
\subsection{CHSH Self-testing threshold}
The CHSH self-testing threshold is given by

\begin{eqnarray}
\sup_{\rho_{AB}}\ &&\int_{\alpha,\beta}\!\!\! d\alpha d\beta \-\ p(\alpha,\beta) \tr (W_{AB}^{\alpha,\beta} \rho_{AB}^{\alpha,\beta})\\
\textnormal{s.t.}\ &&\,\Xi[\rho_{AB}] \leq \frac{1}{2}. \nonumber
\end{eqnarray}
Given the complexity of the expression, its direct maximization seems beyond reach. However, a lower bound on this threshold can be obtained by constructing a state with an extractability of 1/2 and a CHSH value higher than the local bound. We present such an example in the following section.


\section{Lower bound on the CHSH self-testing threshold}
\label{CHSH_Selftesting_threshold}
\subsection{Choice of measurements and state}
\begin{figure}
    \centering
    \includegraphics[width=8cm]{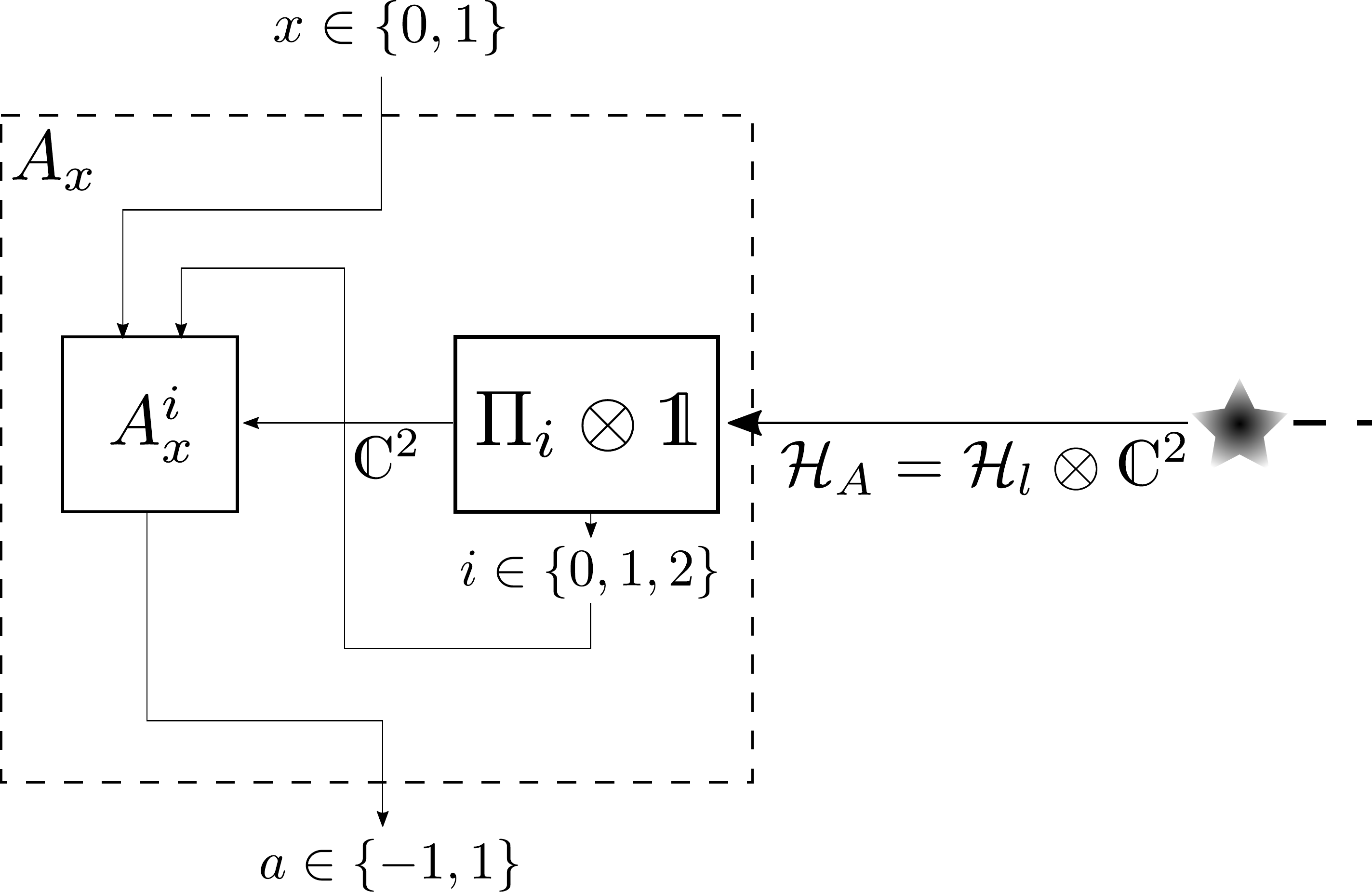}
    \caption{Model of Alice's measurement in the Bell scenario we study. Alice inputs her choice of measurement $x$. The corresponding observable $A_x$ can be seen as a combination of a projector on a classical register and a qubit measurement. The result of the projection $\Pi_i$ is a classical bit $i$ which is attached to a specific qubit measurement $A_x^i$. The outcome $a$ of the qubit measurement is forwarded outside the measurement box, i.e. is accessible to Alice.}
    \label{fig:scenario}
\end{figure}

Following Ref.~\cite{Coopmans2019}, we consider the case where $A_x$ and $B_y$ are made with 3 blocks each that are parametrized by
$\alpha_i = i \pi / 2$ for $i \in \{0,1,2\}$ and similarly for $\beta_j.$ Introducing the compact notation $A_x^i = A_x(\alpha_i)$ and $B_y^j = B_y(\beta_j)$, Eq.~\eqref{eq: jordan} implies that
\be
\label{measurement_choice}
\begin{split}
    A_0^i &= Z \qquad A_1^i \in \{Z, X, -Z\}_{i=0}^2\\
    B_0^j &= H_+ \qquad B_1^j \in \{H_+,H_-,-H_+\}_{j=0}^2.
\end{split}
\ee
For convenience, we decompose the Hilbert spaces $\cH_{A(B)} = \cH_l\otimes\mathds{C}^2$ as a tensor product between the 3-dimensional Hilbert space that carries the label of the block $\ket{i} \in \cH_l$ $(\ket{j}\in \cH_l)$ for $i,j=0,1,2$ and a qubit space $\mathds{C}^2$ on which the measurement operators act as $A_x^i$ and $B_y^j$. The measurement operators of Alice (same holds for Bob) then take the form
\be\label{eq: meas. decomposition}
A_x = \sum_{i=0}^2 \proj{i}\otimes A_x^i.
\ee
as sketched in Fig.~\ref{fig:scenario}. This decomposition of measurements implies the following form for the CHSH operator,
\be
    W_{AB} = \sum_{i,j=0}^2  \ketbra{i}_A \otimes \ketbra{j}_B \otimes W^{ij}_{AB},
\ee
where the expression of each two-qubit observable $W^{ij}_{AB}$ is given in Table~\ref{table:QubitBellObservable}.

\begin{table}[H]
\begin{center}
\begin{tabular}{ c| c c c }
    $W^{ij}_{AB}$ & $j=0$ & $j=1$ & $j=2$ \\
  \hline
  $i=0$ & $2Z\otimes H_+$ & $2Z\otimes H_+$ & $2Z\otimes H_+$ \\
  $i=1$ & $2Z\otimes H_+$ &$\sqrt{2}(X\otimes X+Z\otimes Z)$ & $2X\otimes H_+$ \\  
  $i=2$ & $2Z\otimes H_+$ &$2Z\otimes H_-$ & $-2Z\otimes H_+$    
\end{tabular}
\caption{\label{table:QubitBellObservable}Explicit expression of each two-qubit observable involved in the CHSH operator.}
\end{center}
\end{table}

We choose a state with a similar structure to the one of the Bell operator, that is

\be
\begin{split}
\label{Eq_CHSH2}
    \rho_{AB} &= \nu\ketbra{1}\otimes\ketbra{1}\otimes\rho_{AB}^{11}  \\
    &+(1-\nu)\sum_{\substack{i,j=0 \\ \text{\textbackslash}\{1,1\}}}^2 p_{ij} \ketbra{i} \otimes \ketbra{j} \otimes \rho_{AB}^{ij},
\end{split}
\ee
where $\nu\in[0,1]$ and $p_{ij}$ are coefficients summing up to 1. Here, the sum runs on all pairs $(i,j)$, except $(1,1)$ and $\rho_{AB}^{ij}$ are two-qubit states of the form
\be
    \label{eq:structstate}
    \begin{split}
    \rho_{AB}^{11}&=\ketbra{\Phi^+} \\
    \rho_{AB}^{21}&= \frac{1}{4}(\id\otimes \id +Z\otimes H_-)\\
    \rho_{AB}^{22}&= \frac{1}{4}(\id-Z)\otimes(\id+H_+)\\
    \rho_{AB}^{00}=\rho_{AB}^{02}=\rho_{AB}^{20} &= \frac{1}{4}(\id+Z)\otimes(\id+H_+),
    \end{split}
\ee
with $p_{ij}=0$ for all other values of $i$ and $j$. 
Any such state $\rho_{AB}^{ij}$ $\forall \, \{i,j\}\neq\{1,1\}$ has a contribution to the CHSH score $\tr{\left(W^{ij}_{AB} \rho_{AB}^{ij}\right)}=2.$ By construction, the choice of measurements~\eqref{measurement_choice} and state~\eqref{Eq_CHSH2} thus leads to a CHSH score $S = 2+(2\sqrt{2}-2)\nu$ which is independent of the choice of coefficients $p_{ij}$. 
\subsection{Parametrization of a single-qubit channel} The quantity of interest being the extractabilty of the qubit state $\rho_{AB}$, we need a parametrisation of local maps. As we are interested in the average fidelity which is linear in all channels involved, we can always assume all the channels to be extremal. Single-qubit CPTP maps have been studied in depth by \textit{Verstraete and Verschelde} in Ref.~\cite{Verstraete2002}. It was shown that extremal single-qubit CPTP maps are either unitary transformations or rank 2 maps characterized by the following Kraus operators
\begin{align}\label{eq: Kraus rep}
    K_0&= U\left(\begin{array}{cc}
        s_0 &\\
        & s_1
        \end{array}\right)V^\dag \\
    K_1&= U\left(\begin{array}{cc}
        &\sqrt{1-s_1^2}\\
        \sqrt{1-s_0^2}&
        \end{array}\right)V^\dag
\end{align}
where $s_0,s_1 \in [0,1],$ $U$ and $V$ are unitaries. The unitary transformations can be obtained by setting $s_0= s_1 =1$.

The Bloch vector representation $\bf v$ of a qubit density matrix $\rho= \frac{1}{2} (\id+\bm{\sigma}^\dag {\bf v}))$ is defined  by it's expansion in the Pauli oparators ${\bm \sigma} = (X\, Y\, Z)$. Accordingly, any CPTP map can be represented as an affine transformation of the Bloch ball. Concretely, the affine representation of a CPTP map $\Lambda$ is given by a vector {\bf a} and a matrix $M$ such that for any $\bf{v}$
\be
\Lambda[\frac{1}{2}(\id+\bm{\sigma}^\dag {\bf v})] = \frac{1}{2}\left(\id + {\bm \sigma}^\dag \left({\bf a} + M {\bf v}\right)\right).
\ee
\begin{widetext}
Form the Kraus representation of an extremal rank 2 maps in Eq. \eqref{eq: Kraus rep}, it is straightforward to find its affine representation
    \begin{align}\label{eq: affine rep}
        {\bf a} &= R_U 
            \left(\begin{array}{c}
            0\\
            0\\
            s_0^2-s_1^2
            \end{array}\right)\quad
        M =R_U\left(\begin{array}{ccc}
            s_0s_1 + \sqrt{(1-s_0^2)(1-s_1^2)}&&\\
            &s_0s_1 - \sqrt{(1-s_0^2)(1-s_1^2)}&\\
            && -1 + s_0^2 + s_1^2
            \end{array}\right)R_V^T
    \end{align}
\end{widetext}
where $R_U$ and $R_V$ are rotations in $SO(3),$ cf. \App{app:affrep}. In the following, the affine representation of Alice's (resp. Bob's) map  $\Lambda^i_A$ (resp. $\Lambda^j_B$) performed for register $i$ (resp. $j$) is denoted by the pair $({\bf a}_i, M^i_A)$ (resp. $({\bf b}_j, M^j_B)$). 
\subsection{Relaxation of single-qubit maps} 
\label{sec: relaxation} Let us focus on the registers on \textit{corners}, i.e. $i,j\neq1$. First, we remark that the states $\rho^{ij}_{AB}$ on these blocks only involve a single traceless operator $Z$ ($H_+$) on Alice's side (Bob's side). Therefore, from the matrix part $M^i_A$ of the affine representation of the map $\Lambda^i_A,$ only the vector
\be
 {\bm \zeta}_i = M_A^i \hat {\bf z}
\ee
is relevant for the extracted fidelity. We thus want to express the completely positive (CP) constraint in term of $({\bf a}_i, {\bm \zeta}_i)$ directly. To do so we notice that by definition of the operator norm
\be
|{\bm \zeta}_i| \leq ||M_A^i||= s_0s_1 + \sqrt{(1-s_0^2)(1-s_1^2)},
\ee
where the last expression is the largest singular value and the operator norm of $M_A^i$, as parametrized by $s_0,s_1\in[0,1]$ accordingly to Eq.~\eqref{eq: affine rep}. Using $s_0 = \cos(\widetilde{a_0})$ and $s_1= \cos(\widetilde{a_1})$, we get from the same equation
\begin{align}
\begin{split}
&|{\bf a}_i|^2 + |{\bm \zeta}_i|^2\\
&\leq (s_0^2-s_1^2)^2 + (s_0s_1 + \sqrt{(1-s_0^2)(1-s_1^2)})^2\\
& = (\cos^2(\widetilde{a_0})-\cos^2(\tilde a_1))^2 +\cos^2(\tilde a_0 -\tilde a_1)\\
& = 1- \cos^2(\widetilde{a_0}+\widetilde{a_1})\sin^2(\widetilde{ a_0}-\widetilde{a_1}).
\end{split}
\end{align}
It follows that
\be
a_i^2+  |{\bm \zeta}_i|^2\leq 1\label{eq: const affine}
\ee
with $|{\bf a}_i| = a_i.$ Obviously, the same holds for the map $\Lambda_j^B,$ i.e. 
\be
b_j^2+  |{\bm \eta}_j|^2\leq 1\label{eq: const affine2}
\ee
where ${\bm \eta}_j = M_B^j \hat {\bf h}.$
Hence the maximization over the maps $({\bf a}_i,M_A^i)$ and $({\bf b}_j,M_B^j)$ boils down to a search through $({\bf a}_i,{\bm \zeta}_i)$ and  $({\bf b}_j, {\bm \eta}_j)$ with the constraints of Eqs.~\eqref{eq: const affine}-\eqref{eq: const affine2}.


\subsection{Construction of the example} We shall now construct an explicit example of a state of the form~\eqref{Eq_CHSH2}-\eqref{eq:structstate} leading to a CHSH score $>2$ while having an extractability of at most $\frac{1}{2}$. Such a construction seems challenging at first sight since the extractability is defined from a maximization over all maps, but the previous parametrization of one qubit channels helps significantly simplifying this maximization. Concretely, we consider states $\rho_{AB}$ that are mixtures between the singlet state, the four \textit{corners} states $\rho_{AB}^{ij}$ with $i,j=0$ or $2$ and $\rho_{AB}^{21}.$ We bound the fidelity with respect to the singlet of such states from the parameters of maps $\Lambda_A^1,$ $\Lambda_A^2$ and $\Lambda_B^1$ only. For fixed weights of components of $\rho_{AB},$ we maximize the fidelity over these parameters. We vary the weights until we find the highest Bell score with a state having an extractibility upper bounded by 1/2. \\

We start by fixing the weights of specific corner states to
\be \label{eq: weights}\begin{split}
p_{00}&=p_{02}= \frac{1}{2}p_C q,\\
p_{20}&=p_{22}= \frac{1}{2}p_C (1-q)\\
p_{21}&= 1-p_C
\end{split}
\ee
where $p_C,q\in [0,1].$ Using the action of the maps
\begin{align}
\begin{split}
\Lambda^{i}_A: \frac{1}{2}(\id \pm Z) &\mapsto \frac{1}{2}(\id +{\bm \sigma}^\dag({\bf a}_i \pm {\bm \zeta}_i))\\
\Lambda^{j}_B: \frac{1}{2}(\id + H) &\mapsto \frac{1}{2}(\id +{\bm \sigma}^\dag({\bf b}_j +{\bm \eta}_j))
\end{split}
\label{eq: Lambda ze}
\end{align}
and the relaxation presented above, we can bound the fidelity with respect to the singlet of the state $\rho_c=1/p_C\sum_{i,j=0,2} p_{ij} \ketbra{i}\otimes \ketbra{j}\otimes \rho_{AB}^{ij}$ with $p_C=\sum_{i,j=0,2} p_{ij}.$ In particular, we show in \App{app:cornerfid} that $F_C \leq \frac{1}{4}(1+\epsilon_C)$ where
\begin{equation}
\epsilon_C = \sqrt{(q + (1-q) a_2)^2+(1-q)^2|{\bm \zeta}_2|^2}.
\end{equation}
In the same appendix we also prove that setting $q=\frac 1 2$ is the most constraining case for the singlet extractability. The previous inequality only depends on the parameters of $\Lambda_A^2.$ For example, this implies that unital maps $(a_2=0)$ can at best give a fidelity $F_C =\frac{2+\sqrt{2}}{8} < \frac{1}{2}.$ 

In \App{app:21fid}, we consider $\rho_{AB}^{21}$
and show that its fidelity after extraction $F_{21}\leq \frac{1}{4}(1+\epsilon_{21})$ is such that 
\begin{equation}
    \epsilon_{21}= a_2 b_1 + |{\bm \zeta}_2|\, \sigma_\text{max}(M_B^1).
\end{equation}

Finally, \App{app:phifid} shows that the singlet fidelity after extraction of $\ket{\Phi^+}$ satisfies$F_{\Phi^+} \leq \frac{1}{4}(1+\epsilon_{\Phi^+})$ where
\be
  \epsilon_{\Phi^+} = a_1 b_1 + {\bm \Sigma}(M_A^1)^T  {\bm \Sigma}(M_B^1).
\ee
By combining these bounds, we find that the state $\rho_{AB}$ given in Eqs.\eqref{Eq_CHSH2} and \eqref{eq:structstate} with the weights $p_{ij}$ of Eq.~\eqref{eq: weights} has a fidelity $F_\rho \leq (1+\epsilon_\rho)/4,$ where 
\be
\epsilon_\rho = \nu \epsilon_{\Phi^+} + (1-\nu) p_C \epsilon_{C} + (1-\nu) (1-p_C) \epsilon_{21}
\ee
is a function of only 5 parameters for any fixed values of $p_C$ and $\nu$ (cf. \App{app:fidblock}). For each value of $\nu,$ $\epsilon_\rho$ is maximized over these 5 parameters and then minimized over $p_C$ (cf. \alg{alg:opt}).

\begin{algorithm}[H]
    \caption{Optimization for a self-testing threshold.\label{alg:opt}}
    \begin{algorithmic}
    \State Initialize $\s_{\nu} \gets$ step between violation.
    \For{$\nu\leftarrow 0$ to $1$ by $s_{\nu}$}
        \State $\epsilon_\text{max} = \min\limits_{p_C} ( \max( \epsilon_\rho) ) $
        \If{$\epsilon_\text{max} > 1$ }
            \State return $p_C$ of previous step.
        \EndIf
    \EndFor
    \end{algorithmic}
\end{algorithm}
By doing so, we found a state $\rho$ parametrized by
\begin{equation}
    \label{eq:param}
    \nu=0.061, \quad p_C = 0.61381508,
\end{equation}
the CHSH score corresponding to $\nu=0.061$ being $\approx 2.05.$
\subsection{Certification} $\epsilon_\rho$ being a non-convex function, we need a certification of the result found by the algorithm presented in \alg{alg:opt}. This means that we need to certify that the extractability of the state $\rho_{AB}$ with the parameters given in Eq. \eqref{eq:param} is smaller than or equal to 1/2 or, equivalently, that $\epsilon_{\rho} \leq 1$ for such parameters. Such a certification is obtained by first dividing the compact space of parameters (dimension n=5) in small hypercubes with edges of size $\delta.$ We evaluate the value of $\epsilon_\rho({\bm \varsigma})$ at the central point of each hypercube here parametrized by the vector ${\bm \varsigma}.$ 
In addition, we derive an analytical bound on the norm of the gradient 
\be\nonumber
|\nabla \epsilon_\rho|\leq \iota_{\sup}
\ee
on the whole of its domain. 
It follows that the function $\epsilon_{\rho}$ cannot exceed $1$ within the hypercube centered at ${\bm \varsigma}$
if 
\be
\label{cond_hypercube}
\epsilon_\rho({\bm \varsigma}) \leq 1 -\frac{\sqrt{n}\delta}{2} \iota_{\sup}.
\ee
We eliminate all the hypercubes satisfying this condition. Each hypercube which does not satisfy Eq.~\eqref{cond_hypercube} is divided in $2^n$ smaller hypercubes with edge size $\delta/2$, whose centers are shifted form ${\bm \varsigma}$ by a length $\pm \sqrt{n}\delta/4$ along all diagonals. We then check the condition \eqref{cond_hypercube} with $\delta$ replaced by $\delta/2$ for all the  $2^n$ new hypercubes.
The method we just described is repeated several times and converges for all hypercubes where $\epsilon_\rho$ is below $1$ with a finite gap. After a few iterations, we can exclude all but a small region around the maximum found in \alg{alg:opt}. This maximum corresponds to the full amplitude damping maps -- Alice and Bob simply discard their inputs and prepare a pure product state with $\epsilon(\rho)=1$.  In App.~\ref{sec:analytical}, we prove analytically that $\epsilon_\rho\leq1$ in this small region.
This completes the proof that the singlet extractability of the state $\rho_{AB}$ for the parameter values of Eq.~\eqref{eq:param} is upper bounded by $1/2$. Hence, we have shown that the CHSH self-testing threshold is at least equal to $ 2.05.$\\

\section{Discussion and Conclusion}
\label{Discussion}

Let us come back to our original question: what is the minimal CHSH score ensuring that a nontrivial fraction of $\ket{\Phi^+}$ can be obtained from the state $\rho_{AB}$? In this paper we addressed this question from the self-testing perspective -- Alice and Bob make no assumptions about the source and the measurements boxes, and are allowed to extract the reference state by means of local operations (LO). But this question can also be addressed from different perspectives.

We first consider the situation where Alice and Bob are willing to make some assumptions about the functioning of their devices. In the usual tomographic setting, for example, Alice and Bob have an exact model of their measurements. They trust that those are given by orthogonal Pauli operators described by Eq.~\eqref{eq: jordan} for $\alpha=\beta=\pi/2$. The CHSH score is then given by
\begin{eqnarray}
\nonumber
S &=& \tr\,  \left(\rho_{AB}\, \sqrt{2}(Z\otimes Z +X\otimes X)\right) \\
\nonumber
&\leq& 2 \sqrt{2} \bra{\Phi^+} \rho_{AB} \ket{\Phi^+}.
\end{eqnarray}
A nontrivial fidelity with the reference state can be guaranteed as soon as $S>\sqrt{2}$.

In a semi-device-independent setting where Alice and Bob have no model of the measurement boxes but are ready to trust that the source prepares an unknown two-qubit state $\rho_{AB}\in \cL(\mathds{C}^2\otimes \mathds{C}^2),$ it has been shown that $\rho_{AB}$ has a nontrivial fidelity with a maximally entangled two-qubit state state as soon as the CHSH score exceeds the local bound $S>2$ \cite{Bardyn2009}.

In a fully device-independent framework, one can still ask if the requirement on the minimum CHSH score is lower if Alice and Bob could do more than local operations to extract the reference state. The first step in this direction is to allow them to have access to shared randomness (LOSR). However, since our figure of merit is linear with respect to the extraction strategy of Alice and Bob, shared randomness can not improve the overlap with the reference state. The second step is to allow Alice and Bob to share classical communication (LOCC). Under LOCC, a nontrivial fidelity with the singlet can again be obtained as soon as the local bound is exceeded $S>2$\cite{Bardyn2009}. This can be easily understood from the block-diagonal decomposition of Eq.~\eqref{eq: block proj} obtained from Jordan's lemma. If Alice and Bob make local projections onto their respective orthogonal blocks and exchange the obtained outcomes by means of classical communications, they essentially reduce the problem to the semi-device-independent case. Importantly, by allowing Alice and Bob to exchange classical communication, one abandons the idea of identification up to passive local transformations, which is the spirit of self-testing as we understand it.

In summary, the context of self-testing with CHSH is really special as the self-testing threshold is higher than the local bound. Using a simple parametrisation of extremal single-qubit CPTP maps, we succeeded to bound the extractibility of a whole class of states.  We found a two-qudit state with local dimension $d=6$ that has a trivial extractability but leads to a CHSH score of $\approx 2.05.$ Together with the results presented in Ref. \cite{Jed2016}, this shows that the lowest CHSH score required to self-test the singlet is in the interval $[2.05,2.11].$ Time will tell what is exactly the CHSH self-testing threshold but we now know that it is significantly larger than the local bound.\\

\begin{acknowledgments}
    We are thankful for useful discussion with J\k{e}drzej Kaniewski. We acknowledge funding by the Swiss National Science Foundation (SNSF), through the Grant PP00P2-179109 and 200021-175527 and by the Army Research Laboratory Center for Distributed Quantum Information via the project SciNet. 
\end{acknowledgments}


\bibliographystyle{plainnat}{}
\bibliography{references}


\appendix
\section{From Kraus operator to affine representation of single-qubit CPTP maps}
\label{app:affrep}

In this section we present a way to recover the affine representation of a CPTP map $\Lambda$ -- the pair $({\bf a},M)$ -- given the Kraus operators characterizing this map 
\begin{align}
    K_0&= U\left(\begin{array}{cc}
        s_0 &\\
        & s_1
        \end{array}\right)V^\dag \\
    K_1&= U\left(\begin{array}{cc}
        &\sqrt{1-s_1^2}\\
        \sqrt{1-s_0^2}&
        \end{array}\right)V^\dag
\end{align}
where $s_0,s_1 \in [0,1],$ $U$ and $V$ are unitaries.
\\

We can determine the translation part of $\Lambda$ by applying it on the identity, i.e.
\begin{eqnarray}
\nonumber
   \Lambda [\id] & = & \sum_{i=0}^{1}K_i \id K_i^\dag \\
   \nonumber
    & = & U \left ( \begin{array}{cc}
    1 + (s_0^2 -s_1^2)  & 0 \\
    0 & 1- (s_0^2 -s_1^2)
    \end{array} \right) U^\dag \\
    \nonumber
    &=& \id + {\bm \sigma}^\dag{\bf a}
\end{eqnarray}
where the vector ${\bf a}$ is defined by

\begin{align}
    {\bf a} &= R_U \left(\begin{array}{c}
        0\\
        0\\
        s_0^2-s_1^2
    \end{array}\right).
\end{align}
$R_U$ is the Bloch representation of $U.$ In order to find the matrix $M$, we first study the action of $\Lambda$ on $\tau_x=V \frac{1}{2}\left(\id+X\right) V^\dag$
\begin{align}
\begin{split}
    &\Lambda[\tau_x] = \sum_{i=0}^{1}K_i \tau_x K_i^\dag \\
    \nonumber
      &=  \frac{1}{2} U  \left(\id + (s_0^2-s_1^2)Z + \left(s_0s_1+\sqrt{1-s_0^2}\sqrt{1-s_1^2}\right)X  \right) U^\dag.
      \end{split}
\end{align}
The request that the previous equality can be written as
\begin{equation}
\Lambda[\tau_x]=\frac{1}{2}\left(\id+{\bm \sigma}^\dag({\bf a}+M R_V  \left(\begin{array}{c}
        1  \\
        0  \\
        0
    \end{array}\right)\right)
\end{equation}
fixes some elements of $M$ 
\begin{equation}
M =  R_U \left(\begin{array}{ccc}
        s_0s_1+\sqrt{1-s_0^2}\sqrt{1-s_1^2} & . & . \\
        0 & . & . \\
        0 & . & .
    \end{array}\right) R_V^\dag
\end{equation}
The missing elements can be found in a similar way by applying $\Lambda$ to $V \frac{1}{2}\left(\id+Y\right) V^\dag$ and $V \frac{1}{2}\left(\id+Z\right) V^\dag.$
This characterizes the pair $({\bf a},M)$ up to unitaries (6 remaining degrees of freedom).


\section{Bounding the fidelity}
In this section, we show how to bound the singlet fidelity after extraction of the state
\begin{widetext}
\be
\begin{split}
    \rho_{AB} &= \nu\ketbra{1}\otimes\ketbra{1}\otimes\rho_{AB}^{11}  \\
    &+(1-\nu)\sum_{\substack{i,j=0 \\ \text{\textbackslash}\{1,1\}}}^2 p_{ij} \ketbra{i} \otimes \ketbra{j} \otimes \rho_{AB}^{ij},
\end{split}
\ee
where the expression of each component is given by
\begin{center}
    \begin{tabular}{ c| c c c }
        \hline
        \multicolumn{4}{c}{$\rho_{AB}^{ij}$} \\
    \hline
        \diagbox{i}{j}  & $0$ & $1$ & $2$ \\
    \hline
    $0$ & $\ketbra{z,h_+}$ &  & $\ketbra{z,h_+}$ \\
    $1$ &  &$\ketbra{\Phi^+}$ &  \\  
    $2$ & $\ketbra{z,h_+}$ & $\frac{1}{4}(\id \otimes \id +Z\otimes H_-)$ & $\ketbra{-z,h_+}$   
\end{tabular}
\end{center}
\end{widetext}

where the corner states can be conveniently expressed as $\ketbra{\pm z,h_+} = \frac{1}{4}(\id \pm Z)\otimes(\id + H_+)$.  The wights  $p_{ij}$ are given by
\begin{center}
\begin{tabular}{ c| c c c }
    \hline
        \multicolumn{4}{c}{$p_{ij}$} \\
        \hline
        \diagbox{i}{j}  & $0$ & $1$ & $2$ \\
 \hline
 $0$ & $\frac{q p_C}{2}$ & $0$ & $\frac{q p_C}{2} $ \\
 $1$ & $0$ &  & $0$ \\  
 $2$ & $\frac{(1-q) p_C}{2}$ &$(1-p_C)$ & $\frac{(1-q) p_C}{2}$    
\end{tabular}
\end{center}
in accordance with the main text. To bound the singlet fidelity of $\rho_{AB},$ we proceed component by component. 

\subsection{\textit{Corner} states}
\label{app:cornerfid}

First, let us note that from the Pauli expansion of the state 
\be
\ketbra{\Phi^+}= \frac{1}{4}(\id \otimes \id + X\otimes X -Y \otimes Y + Z\otimes Z)
\ee
it follows that 
\begin{align}
    & F\Big(\frac{1}{4}(\id +{\bm \sigma}^\dag {\bf a})\otimes(\id + {\bm \sigma}^\dag {\bf b}), \ketbra{\Phi^+}) \Big) \nonumber\\
    = & \frac{1}{4}(1 + {\bf a}^T \,J\, {\bf b}) =\frac{1}{4}(1 + {\bf a}^T {\bf b}')
    \label{eq: F prod}
\end{align}
where 
\begin{align}
    \label{eq:J}
    \begin{split}
     &J =  
    \left(\begin{array}{c c c}
    1&&\\
    &-1&\\
    &&1\\
    \end{array}\right)\\
    & \text{and} \quad {\bf b}'=  J {\bf b} .
    \end{split}
\end{align}

With this expression and the notation ${\bf a_i}, {\bm \zeta}_i, {\bf b}_j$, and ${\bm \eta}_j$ of the main text \ref{sec: relaxation} one concludes that the singlet extractability for the mixture of the corner states
\be\rho_C= \frac{1}{p_C}\sum_{i,j=0,2} p_{ij} \ketbra{i} \otimes \ketbra{j} \otimes \rho_{AB}^{ij}
\ee
is given by
\be
F_C =\frac{1}{4}(1+\bar \epsilon_C)
\ee
with
\begin{align}\nonumber
    &\bar \epsilon_C =
    \frac{q}{2} ({\bf a}_0+ {\bm \zeta}_0)^T({\bf b}'_0+ {\bm \zeta}'_0) +\frac{q}{2} ({\bf a}_0+ {\bm \zeta}_0)^T({\bf b}'_2+ {\bm \zeta}'_2) +\\
    & \frac{1-q}{2} ({\bf a}_2+ {\bm \zeta}_2)^T({\bf b}'_0+ {\bm \zeta}'_0) + \frac{1-q}{2} ({\bf a}_2- {\bm \zeta}_2)^T({\bf b}'_2+ {\bm \zeta}'_2) \nonumber
\end{align}
Introducing the quantities ${\bf A}_0 = {\bf a}_0 + {\bm \zeta}_0$ and ${\bf B}_j= {\bf b}_j' +  {\bm \eta}_j'$, that satisfy $|{\bf A}_0|,|{\bf B}_j|\leq 1$, the expression above simplifies to
\begin{align}
    \bar \epsilon_C &=
    \frac{1}{2}(q{\bf A}_0 + (1-q){\bf a}_2)^T ({\bf B}_0 + {\bf B}_2) \\
    &+ \frac{1}{2}(1-q) {\bm \zeta}_2 ({\bf B}_0 - {\bf B}_2).
\end{align}
Since we are interested in a convex optimisation over extremal maps, we can without loss of generality fix the norms of the vectors $|{\bf A}_0|,|{\bf B}_j| =  1$ to their maximum values. 

The sum and difference of arbitrary unit vectors can be equivalently parametrized as
\begin{align}
    {\bf B}_0 + {\bf B}_2 &= 2s_\theta {\bf C} \\
     {\bf B}_0 - {\bf B}_2 &= 2 c_\theta {\bf C}_\perp
\end{align}
with ${\bf C}$ and ${\bf C}_\perp$ unit vectors satisfying ${\bf C}^T{\bf C}_\perp =0$. Hence
\begin{equation}
\bar \epsilon_C \leq  s_\theta |q{\bf A}_0 + (1-q){\bf a}_2|  +  c_\theta (1-q)|{\bm \zeta}_2|.
\end{equation}
Using $A_0=1$ we obtain,
\begin{equation}
 \bar \epsilon_C \leq  s_\theta (q + (1-q) a_2)  +  c_\theta (1-q)|{\bm \zeta}_2|.
\end{equation}
Finally, thanks to the inequality $|a\cos(x) + b\sin(x)|\leq\sqrt{a^2+b^2}$, we have 
\begin{equation}
F_C \leq \frac{1}{4}(1+\epsilon_C)
\end{equation}
with
\begin{equation}
\epsilon_C =  \sqrt{(q + (1-q) a_2)^2+(1-q)^2|{\bm \zeta}_2|^2}.
\end{equation}
\subsection{$(2,1)$ block}
\label{app:21fid}

For the $(2,1)$ block analogously to Eq.~\eqref{eq: F prod} we express the fidelity as
\begin{equation}
    F_{21} = \frac{1}{4}\left(1+ {\bf a}_2^T {\bf b}_1' + {\bm \zeta}_2^T {\bf m}_1' \right),
\end{equation}
where ${\bf m}_1$ is the Bloch vector representation of $\Lambda_B^1[H_-]$. \\
We define $a_2=|{\bf a}_2|$ and $b_1=|{\bf b}_1'|$. The fidelity is now bounded by
\begin{equation}
    F_{21} \leq  \frac{1}{4}(1+a_2 b_1 + |{\bm \zeta}_2||{\bf m}_1'|).
\end{equation}
Let $\sigma_\text{max}(M_B^1)$ be the maximal singular value of $M_B^1.$ Since $|{\bf m}_1'|\leq \sigma_\text{max}(M_B^1),$ we have 
\begin{equation}
     F_{21} \leq  \frac{1}{4}(1+ \epsilon_{21})
\end{equation}
with
\begin{equation}
    \epsilon_{21} =  a_2 b_1 + |{\bm \zeta}_2|\, \sigma_\text{max}(M_B^1).
\end{equation}
\subsection{$\Phi^+$ state}
\label{app:phifid}

The fidelity of the singlet after extraction is a function of the two \textit{central} maps parametrized by $({\bf a}_1, M_A^1)$ and $({\bf b}_1, M_B^1)$
\begin{equation}
    F_{\Phi^+} = \frac{1}{4}(1+ {\bf a}_1^T {\bf b}'_1 + \tr J M_A^1 J M_B^1),
\end{equation}
where $J$ is the matrix defined in \eq{eq:J}. Using the fact that $J M J$ has the same singular values as $M$, and the relation shown in Ref. ~\cite{Mirsky1975}, 
\begin{equation}
    \tr M M' \leq {\bm \Sigma}(M)^T  {\bm \Sigma}(M'),
\end{equation}
where ${\bm \Sigma}(M)$ is the ordered vector of singular values of $M,$ we get 
\begin{equation}
    F_{\Phi^+} \leq  \frac{1}{4}(1+\epsilon_{\Phi^+})
\end{equation}
with
\begin{equation}
    \epsilon_{\Phi^+} = ( a_1 b_1 + {\bm \Sigma}(M_A^1)^T  {\bm \Sigma}(M_B^1)).
\end{equation}

\subsection{Singlet fidelity over all blocks}
\label{app:fidblock}
Combing the bounds derived on the last three sections, we obtain that the fidelity of the overall state $\rho_{AB}$ is upper-bounded by
\begin{equation}     
     F_\rho\leq \frac{1}{4}(1+\bar \epsilon)
\end{equation}
with
\begin{eqnarray}
    \bar \epsilon &=& (1-\nu)(p_C \epsilon_C + (1-p_C)\epsilon_{21}) + \nu\epsilon_{\Phi^+}\\
    \label{eq:convsqrt}
    &=& \nu( a_1 b_1 + {\bm \Sigma}(M_A^1)^T  {\bm \Sigma}(M_B^1)) \\
    \nonumber
    &&+ (1-\nu) p_C \sqrt{(q + (1-q) a_2)^2+(1-q)^2|{\bm \zeta}_2|^2} \\
    \nonumber
    &&+  (1-\nu) (1-p_C) (a_2 b_1 + |{\bm \zeta}_2| \, \sigma_\text{max}(M_B^1)).
\end{eqnarray}
First, note that the rhs of~\eq{eq:convsqrt} is always maximized when the variables $a_2$ and $|{\bm \zeta}_2|$ saturate the constraint given in Eq.~\eqref{eq: const affine} $$a_2^2+  |{\bm \zeta}_2|^2 = 1.$$ Hence we can replace them by a single variable $\theta \in [0,\pi/2]$ with
\be
    a_2= \cos(\theta) \quad  |{\bm \zeta}_2|= \sin(\theta),
\ee
and rewrite 
\be
\begin{split}
\bar \epsilon &\leq \nu( a_1 b_1 + {\bm \Sigma}(M_A^1)^T  {\bm \Sigma}(M_B^1)) \\
    &+ (1-\nu) p_C \sqrt{(q + (1-q) \cos\theta)^2+(1-q)^2\sin^2\theta} \\
    &+  (1-\nu) (1-p_C) (\cos(\theta) b_1 + \sin(\theta) \, \sigma_\text{max}(M_B^1)).
\end{split}
\ee
The only term involving $q$ in the inequality above $q$ reads
\begin{align}
    & \sqrt{(q + (1-q) \cos\theta)^2+(1-q)^2\sin^2\theta} \\
    = & \sqrt{(q^2 + 2q(1-q)\cos\theta + (1-q)^2} \\
    = & \sqrt{1-4 \, q(1-q)\sin^2(\frac{\theta}{2})},\\
\end{align}
which attains its minimum of $\cos(\frac \theta 2)$ for $q=\frac{1}{2}$. Hence, we fix $q=\frac 1 2$ since it is the most constraining case for the extracted fidelity
\be
\begin{split}
\bar \epsilon &\leq \nu( a_1 b_1 + {\bm \Sigma}(M_A^1)^T  {\bm \Sigma}(M_B^1)) \\
    &+ (1-\nu) p_C \,\cos\left(\frac \theta 2\right) \\
    &+  (1-\nu) (1-p_C) (\cos(\theta) b_1 + \sin(\theta) \, \sigma_\text{max}(M_B^1)).
\end{split}
\ee \\

We then look at ${\bm \Sigma}(M_A^1)^T  {\bm \Sigma}(M_B^1).$ The singular values of $M$ which is given in Eq.~\eqref{eq: affine rep} are
\be
\begin{split}
    \lambda_1&=s_0s_1 + \sqrt{(1-s_0^2)(1-s_1^2)}  \\
    \lambda_2&= s_0s_1 - \sqrt{(1-s_0^2)(1-s_1^2)}  \\
    \lambda_3 &= -1 + s_0^2 + s_1^2
\end{split}
\ee
 and are ordered in absolute value as
 \be
  |\lambda_1|\geq |\lambda_2|\geq |\lambda_3|.
 \ee
Furthermore, $\lambda_1$ is positive on the whole domain, and the two other values $\lambda_2$ and $\lambda_3$ always have the same sign. Hence, we can separate $\bar \epsilon$ in two parts,
\begin{align}
    \label{eq:erho}
    \begin{split}
    &\bar \epsilon^\pm \leq \nu( a_1 b_1 + \lambda_1^{A} \lambda_1^{B} \pm (\lambda_2^{A} \lambda_2^{B} + \lambda_3^{A} \lambda_3^{B})) \\
        &+ (1-\nu) p_C \cos\left(\frac \theta 2\right) \\
&+  (1-\nu) (1-p_C) (\cos\theta \, b_1 + \sin\theta \, \lambda_1^{B})=\epsilon_\rho^\pm
    \end{split}
\end{align}
where $\lambda_x^{A(B)}$ is the $x^{\text{th}}$ greatest singular value of $M_A^1(M_B^1)$, parametrized by $s_0^A$ and $s_1^A$ ($s_0^B$ and $s_1^B$) respectively. Finally, from Eq.\eqref{eq: affine rep} we also get
\be
a_1= (s^A_0)^2- (s^A_1)^2 \qquad b_1= (s_0^B)^2- (s_1^B)^2.
\ee
One could ensure that $a_1$ and $b_1$ are positive by requiring $s_0^{A(B)}\geq s_1^{A(B)}$, which is always possible because the matrices $M_1^{A(B)}$ and their singular values $\lambda_i^{A (B)}$ are invariant under the exchange of $s_0^{A(B)}$ and $ s_1^{A(B)}$.

We have shown how the optimization of the extractability of the state $\rho_{AB}$ over all possible local
extraction strategies for Alice and Bob
can be reduced to the maximization of a simple function $\epsilon_\rho^\pm$ -- the right hand side of  Ineq.~\eqref{eq:erho} -- over five parameters $s_0^A,s_1^A, s_0^B, s_1^B,\theta$. This function can be maximized numerically. However, a naive maximization does not guarantee that the value of $\epsilon_\rho$ remains below $1$ on the whole of its domain. In the next appendix, we show how such a guarantee can be obtained.  

\section{Certification}
~\label{app:cert}

In this appendix we describe the certification process behind 
\be
    \max\limits_{s_0^A,s_1^A,s_0^B,s_1^B,\theta} \epsilon_\rho^\pm \leq 1,
\ee
where the parameters $\nu,p_C,q$ are fixed.\\
Our approach is a two-step process. First, we exclude sub-domains of parameters' space using an algorithm which evaluates $\epsilon_\rho^\pm$ on a grid in this subspace and compute the possible maximum evolution of $\epsilon_\rho^\pm$ for hypercubes centered on each grid elements using a bound on maximum values each partial derivative can take. Then, on remaining region, we use some bounds on trigonometric functions to analytically show that $\epsilon_\rho^\pm$ does not exceed one.\\

\subsection{Maxima certification algorithm based on partial derivative}
~\label{app:certalg}
\textit{Bound on partial derivative --} First, we aim to bound the maximum value the partial derivatives of our expression can take. In other words, we want to have quantities
\begin{equation}
    \max\limits_{s_0^A,s_1^A,s_0^B,s_1^B,\theta}\left({\frac{\partial \epsilon_\rho^\pm}{\partial x}}\right)
\end{equation}
for every $x\in\{s_0^A,s_1^A,s_0^B,s_1^B,\theta\}$.
Partial derivatives of $\epsilon_\rho^\pm$ as formulated in \eq{eq:erho}, can lead to infinity when they are maximized over every variable. Such behavior can be circumvented \textit{via} the following change of variables
\begin{align}
    s_0^A \rightarrow \cos(\widetilde{a_0})&,\quad \text{with}\; \widetilde{a_0}\in [ 0,\frac{\pi}{2}], \\
    s_1^A \rightarrow \cos(\widetilde{a_1})&,\quad \text{with}\; \widetilde{a_1}\in [ 0,\frac{\pi}{2}], \\
    s_0^B \rightarrow \cos(\widetilde{b_0})&,\quad \text{with}\; \widetilde{b_0}\in [ 0,\frac{\pi}{2}], \\
    s_1^B \rightarrow \cos(\widetilde{b_1})&,\quad \text{with}\; \widetilde{b_1}\in [ 0,\frac{\pi}{2}].
\end{align}
It is sufficient to explore the case where $\widetilde{a_0} \leq \widetilde{a_1}$ and $\widetilde{b_0}\leq \widetilde{b_1}$ to conserve the symmetries defined in the core paper.
$\epsilon_\rho^\pm$ is now expressed as
\begin{widetext}
\begin{align}
    \begin{split}
    \label{eq:epsan}
        \epsilon_\rho^\pm  (\widetilde{a_0},\widetilde{a_1},\widetilde{b_0},\widetilde{b_1},\theta, \nu , & p_C, q)  = 
    \nu \colvec{\cos ^2(\widetilde{a_0})-\cos ^2(\widetilde{a_1}) \\ \cos(\widetilde{a_0})\cos(\widetilde{a_1})+\sin(\widetilde{a_0})\sin(\widetilde{a_1}) \\ \pm (\cos(\widetilde{a_0})\cos(\widetilde{a_1})-\sin(\widetilde{a_0})\sin(\widetilde{a_1})) \\ \pm(-1+\cos ^2(\widetilde{a_0})+\cos  ^2(\widetilde{a_1}))}\cdot \colvec{\cos ^2(\widetilde{b_0})-\cos ^2(\widetilde{b_1}) \\ \cos(\widetilde{b_0})\cos(\widetilde{b_1})+\sin(\widetilde{b_0})\sin(\widetilde{b_1}) \\ \cos(\widetilde{b_0})\cos(\widetilde{b_1})-\sin(\widetilde{b_0})\sin(\widetilde{b_1}) \\ -1+\cos ^2(\widetilde{b_0})+\cos  ^2(\widetilde{b_1})} \\
    & + (1-\nu) p_C \cos\left(\frac \theta 2\right) \\
    & + (1-\nu) (1-p_C) (\cos(\theta)(\cos ^2(\widetilde{b_0})-\cos ^2(\widetilde{b_1}))   + \sin(\theta)(\cos(\widetilde{b_0})\cos(\widetilde{b_1})+\sin(\widetilde{b_0})\sin(\widetilde{b_1}))). 
    \end{split}
\end{align}
    The partial derivatives of~\eq{eq:epsan} are given by
    \begin{align}
        \label{eq:bnd-test}
        \frac{\partial \epsilon_\rho^\pm}{\partial \widetilde{a_0}} = &\mp \nu  (2 \sin (\widetilde{a_0}) \cos (\widetilde{a_1}) \cos (\widetilde{b_0}) \cos (\widetilde{b_1}) -2 \cos (\widetilde{a_0}) \sin (\widetilde{a_1}) \sin (\widetilde{b_0}) \sin (\widetilde{b_1}) \pm \sin (2 \widetilde{a_0}) \cos (2 \widetilde{b_0})) \\
        \frac{\partial \epsilon_\rho^\pm}{\partial \widetilde{a_1}} =  &\mp \nu  (2 \cos (\widetilde{a_0}) \sin (\widetilde{a_1}) \cos (\widetilde{b_0}) \cos (\widetilde{b_1})-2 \sin (\widetilde{a_0}) \cos (\widetilde{a_1}) \sin (\widetilde{b_0}) \sin (\widetilde{b_1}) \pm \sin (2 \widetilde{a_1}) \cos (2 \widetilde{b_1})) \\
    \end{align}
    \begin{align}
        \begin{split}
            \frac{\partial \epsilon_\rho^\pm}{\partial \widetilde{b_0}} = & \mp \nu  (2 \cos (\widetilde{a_0}) \cos (\widetilde{a_1}) \sin (\widetilde{b_0}) \cos (\widetilde{b_1})-2 \sin (\widetilde{a_0}) \sin (\widetilde{a_1}) \cos (\widetilde{b_0}) \sin (\widetilde{b_1}) \pm \cos (2 \widetilde{a_0}) \sin (2 \widetilde{b_0})) \\
            & +  (\nu -1) (1-p_C) (\sin (\theta ) \sin (\widetilde{b_0}-\widetilde{b_1})+\sin (2 \widetilde{b_0}) \cos (\theta ))
        \end{split} \\
        \begin{split}
            \frac{\partial \epsilon_\rho^\pm}{\partial \widetilde{b_1}} = \,&2 \nu (\pm \sin (\widetilde{a_0}) \sin (\widetilde{a_1}) \sin (\widetilde{b_0}) \cos (\widetilde{b_1})- \sin (\widetilde{b_1}) ( \mp \cos (\widetilde{a_0}) \cos (\widetilde{a_1}) \cos (\widetilde{b_0})+\cos (2 \widetilde{a_1}) \cos (\widetilde{b_1})) \\
            &+ (\nu -1) (p_C-1) (\sin (\theta ) \sin (\widetilde{b_0}-\widetilde{b_1})+\sin (2 \widetilde{b_1}) \cos (\theta ))
        \end{split} \\
        \begin{split}
            \frac{\partial \epsilon_\rho^\pm}{\partial \theta} = \,&  (1-\nu) ( (1-p_C) (\sin (\theta ) \sin (\widetilde{b_0}-\widetilde{b_1}) \sin (\widetilde{b_0}+\widetilde{b_1})+\cos (\theta ) \cos (\widetilde{b_0}-\widetilde{b_1})) - p_C \frac{\sin\left(\frac \theta 2\right)}{2}).
        \end{split}
    \end{align}
    To bound above derivatives, we naively set lowest value to negative terms, and maximum value to positive terms, independently of other terms. As an example,
    \begin{equation}
        \frac{\partial \epsilon_\rho^+}{\partial \widetilde{a_0}} = \nu  (-2 \underbrace{\sin (\widetilde{a_0}) \cos (\widetilde{a_1}) \cos (\widetilde{b_0}) \cos (\widetilde{b_1})}_{\geq 0 }  + 2\underbrace{\cos (\widetilde{a_0}) \sin (\widetilde{a_1}) \sin (\widetilde{b_0}) \sin (\widetilde{b_1})}_{\leq 1} -\underbrace{\sin (2 \widetilde{a_0}) \cos (2 \widetilde{b_0})}_{\geq  -1}) \leq 3\nu.
    \end{equation}

Applying this method for all partial derivative we found bounds,
\begin{align}
    \frac{\partial \epsilon_\rho^\pm}{\partial \widetilde{a_0}},\frac{\partial \epsilon_\rho^\pm}{\partial \widetilde{a_1}} &\leq 3\nu \\
    \frac{\partial \epsilon_\rho^\pm}{\partial \widetilde{b_0}},\frac{\partial \epsilon_\rho^\pm}{\partial \widetilde{b_1}} &\leq 3\nu + (1-p_C)(1-\nu) \\
    \frac{\partial \epsilon_\rho^\pm}{\partial \theta} &\leq (1-p_C)(1-\nu).
\end{align}
Thus, we can obtain a bound on the norm of the gradient of $\epsilon_\rho^\pm$, given by,
\begin{equation}
\label{eq:bgrad}
    |\nabla \epsilon_\rho^\pm| \leq \sqrt{2(3\nu)^2+2(3\nu+(1-p_C)(1-\nu))^2+(1-p_C)(1-\nu)^2}=\iota_{\text{sup}}.
\end{equation}
This bound sets a limit on how fast $\epsilon_\rho^\pm$ can change in any direction. It allows one to bound the value of $\epsilon_\rho^\pm$ on the whole of its domain from a list of values it admits on a finite grid of points.

\end{widetext}

\textit{Certification algorithm --} We built an algorithm to certify that a Lipschitz continuous function -- a multivariate scalar function $f:\mathbb{R}^n \to \mathbb{R}$, with a bounded gradient $|\nabla f|\leq \iota_{\text{sup}}$-- on a close domain of definition $D\in \mathbb{R}^n$ is bounded by some constant $\lambda$\footnote{We provide a Python implementation of our algorithm at \url{https://gitlab.com/plut0n/bcert}}. \\

We start by creating a grid on $D$ of dimension $n$, where every point $p\in\mathbb{R}^n$ are separated by an interval $\Delta=\delta$ in every directions. For each element of the grid $p$ we evaluate $f(p)$. 
Since the function $f$ is Lipschitz continuous, its value on the whole hypercube with edge length $\Delta$ centered at $p$ can not be too different from its value in the center $f(p)$. Precisely it is bounded by
\begin{equation}
    f_{\text{max}}(p,\Delta) = f(p) + \frac{\sqrt{n}\Delta}{2}\iota_{\text{sup}}.
\end{equation}\\
Therefore, $f_\text{max}(p,\Delta)\leq \lambda$ certifies that the function is bounded by $\lambda$ on the whole $\Delta$-hyper around $p$, and we can eliminate all such hypercubes from further consideration.

\begin{figure}[t]
    \includegraphics[width=\columnwidth]{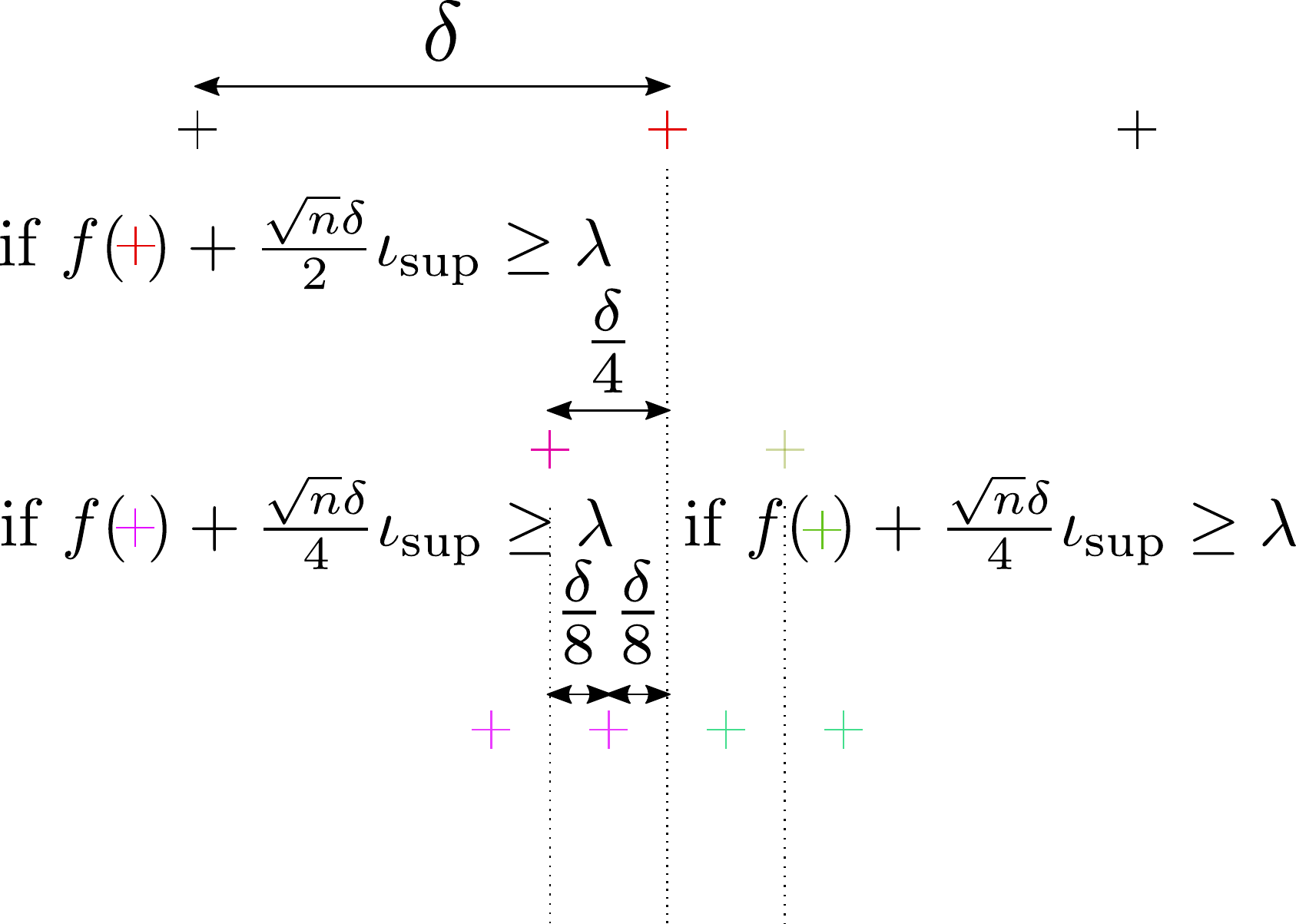}
    \caption{A schematic representation of the recursive part of the algorithm certifying an upperbound $\lambda$ of a $1$-dimensional ($n=1$) scalar function with bounded gradient $\iota_\text{sup}$, on an initial grid interval of $\delta$.}
    \label{fig:algcert}
\end{figure}

For every point with $f_\text{max}(p,\Delta)> \lambda$, we call the algorithm recursively. The domain this time is the hyper-cube centered at $p$, on which we create a smaller grid of step $\Delta=\delta/2$.
 We generate new points $p'$ around $p$ by creating points at a distance $ \frac{\sqrt{n}\Delta}{2}$ of $p$ using,
\begin{equation}
    p' = p + \frac{\sqrt{n}\Delta}{2}d \quad \forall d\in \mathcal{D},
\end{equation}
 where $\mathcal{D}$ is the set of vertices of the $n$-dimensional hypercube centered on 0 and of size 2, i.e. every diagonals of the hypercube. This process might generate some points $p'$ outside of the domain $D$, we thus exclude such generated point. We then repeat the above process for every thus generated new point.
The algorithm converges if there is a finite gap $G$ between the maximum of $f$ on $D$ and the constant
$\lambda$ 
\be
\sup_D f \leq \lambda- G.
\ee
The minimal possible required grid step needed for convergence is then given by $\Delta = \frac{2 G}{\iota_{\text{sup}}\sqrt{n}}$.
Parallelization of such an algorithm can be implemented since every hypercubes are independent of each other. A sketch of the algorithm for a  function $f :\mathbb{R}\to \mathbb{R}$ is depicted in \fig{fig:algcert}. \\

To ensure that the extractability of the state $\rho_{AB}$ we constructed does not exceed $1/2$ we applied the algorithm on the function
\begin{equation}
    f(x,\nu,p_C,\theta) = \max (\epsilon_\rho^+(x,\nu,p_C,q),\epsilon_\rho^-(x,\nu,p_C,q))
\end{equation}
where 
\be
x=(\widetilde{a_0},\widetilde{a_1},\widetilde{b_0},\widetilde{b_1},\theta) \in D= [0,\frac{\pi}{2}]\times\dots\times [0,\frac{\pi}{2}],
\ee
$\epsilon_\rho^\pm$ are the functions defined in \eq{eq:epsan} and $(\nu,p_C,q)$ are fixed to the values given in~\eq{eq:param}. The bound to guarantee is set to $\lambda=1$. We used~\eq{eq:bgrad} for the bound on the gradient. We ran this algorithm on the whole domain of definition $D$ of $f$ with the exception of a small hypercube $C_\text{ex}$ with edge size $\frac{\pi}{16}$ containing the point  $x_\text{exc}=(0,\frac{\pi}{2},0,\frac{\pi}{2},0)$ as a vertex. The region $C_\text{ex}$ is excluded because the function $f(x_\text{exc})=1$ saturates the bound and the algorithm would not converge on its neighbourhood. As explained in the main text $x_\text{exc}$ corresponds to the extraction strategy where Alice and Bob discard $\rho_{AB}$ and prepare a product state whose fidelity with the singled is exactly one half.

Our algorithm did stop on the domain $D \setminus C_\text{ex}$. Hence, we certified that the extracted fidelity of our example state $\rho_{AB}$ never exceeds 1/2 for most  maps. We then need to prove that it is also the case on the remaining hypercube $C_\text{ex}$ of side $\frac{\pi}{16}$ with an extremal vertex $x_\text{exc}$. \\

\subsection{Analytical proof on the remaining subspace}
\label{sec:analytical}

The remaining hypercube $C_\text{ex}$ containing $x_\text{exc}$ is defined by
\begin{align}
    \widetilde{a_0},\widetilde{b_0},\theta & \in [0,\frac{\pi}{16}], \\
    \widetilde{a_1},\widetilde{b_1} & \in [\frac{\pi}{2}-\frac{\pi}{16},\frac{\pi}{2}].
\end{align}
In this section we provide a proof that $\epsilon_\rho^\pm$ achieves its maximum value $1$ in $x_\text{exc}$. \\

\begin{widetext}
The expression of $\epsilon_\rho^\pm$ for $\theta=0$, reads
\begin{align}
\begin{split}
\label{eq:epst0}
    \epsilon_\rho^\pm  =  &
    \nu \colvec{\cos ^2(\widetilde{a_0})-\cos ^2(\widetilde{a_1}) \\ \cos(\widetilde{a_0})\cos(\widetilde{a_1})+\sin(\widetilde{a_0})\sin(\widetilde{a_1}) \\ \pm (\cos(\widetilde{a_0})\cos(\widetilde{a_1})-\sin(\widetilde{a_0})\sin(\widetilde{a_1})) \\ \pm(-1+\cos ^2(\widetilde{a_0})+\cos  ^2(\widetilde{a_1}))}\cdot \colvec{\cos ^2(\widetilde{b_0})-\cos ^2(\widetilde{b_1}) \\ \cos(\widetilde{b_0})\cos(\widetilde{b_1})+\sin(\widetilde{b_0})\sin(\widetilde{b_1}) \\ \cos(\widetilde{b_0})\cos(\widetilde{b_1})-\sin(\widetilde{b_0})\sin(\widetilde{b_1}) \\ -1+\cos ^2(\widetilde{b_0})+\cos  ^2(\widetilde{b_1})} \\
     & + (1-\nu)( p_C\cos(\frac{\theta}{2}) + (1-p_C)(\cos(\theta)(\cos(\widetilde{b_0})^2 -\cos(\widetilde{b_1})^2)) +\sin(\theta) (\cos(\widetilde{a_0})\cos(\widetilde{a_1})+\sin(\widetilde{a_0})\sin(\widetilde{a_1}) ) ).
\end{split}
\end{align}

By introducing a new parametrisation,
\begin{align}
    \mu_a =& \frac{\pi}{2}- (\widetilde{a_1} + \widetilde{a_0})  \quad &\mu_b = \frac{\pi}{2}-(\widetilde{b_1} + \widetilde{b_0}) \\ 
    \delta_a =& \frac{\pi}{2}-(\widetilde{a_1} - \widetilde{a_0}) \quad &\delta_b = \frac{\pi}{2}-(\widetilde{b_1} - \widetilde{b_0} )
\end{align}

with $\delta \in [0,\frac{\pi}{4}]$ and $\mu\in [-\frac{\pi}{8},+\frac{\pi}{8}],$ allows one to rewrite \eq{eq:epst0}  as
\be
\begin{split}
\epsilon^\pm_\rho &=\nu \left(
\begin{array}{c}
 \cos \left(\delta _a\right) \cos \left(\mu _a\right) \\
 \sin \left(\delta _a\right) \\
  \pm \sin \left(\mu _a\right) \\
   \pm \sin \left(\delta _a\right) \sin \left(\mu _a\right) \\
\end{array}
\right)\cdot
\left(
\begin{array}{c}
 \cos \left(\delta _b\right) \cos \left(\mu _b\right) \\
 \sin \left(\delta _b\right) \\
 \sin \left(\mu _b\right) \\
 \sin \left(\delta _b\right) \sin \left(\mu _b\right) \\
\end{array}
\right)\\
&+(1-\nu)( p_C\cos(\frac{\theta}{2}) + (1-p_C)(\cos(\theta)\cos \left(\delta _b\right) \cos \left(\mu _b\right)+\sin(\theta)\sin(\delta_b))).
\end{split}
\ee
As the two branches of the function $\epsilon^\pm_\rho$ can be mapped into each other with $\mu_a\to -\mu_a$ it is sufficient to consider the branch with the plus sign. In addition, since $\epsilon_\rho^+$ is maximized when all the trigonometric functions are positive we can restrict $\mu_{a(b)}\in[0,\frac \pi 8]$ without loss of generality.\\

With the help of the identities
\be
\begin{split}
&\cos \left(\delta _a\right) \cos \left(\mu _a\right) \cos \left(\delta _b\right) \cos \left(\mu _b\right)+ 
\sin \left(\delta _a\right) \sin \left(\mu _a\right) \sin \left(\delta _b\right) \sin \left(\mu _b\right)\\&=
 \frac{1}{2} \cos (\delta_a+\mu_a)\cos(\delta_b+\mu_b) +\frac{1}{2}\cos(\delta_a -\mu_a)\cos(\delta_b-\mu_b) 
\end{split}
\ee
and
\be
\cos(\theta)\cos(\delta_b)\cos(\mu_b) = \frac{\cos(\theta)}{2} \left( \cos(\delta_b - \mu_b) + \cos(\delta_b + \mu_b) \right)
\ee
we rewrite $\epsilon_\rho^+$ as
\be
\begin{split}
\epsilon_\rho^+ &=  \frac \nu 2 \cos (\delta_a+\mu_a)\cos(\delta_b+\mu_b) + \frac \nu 2 \cos (\delta_a-\mu_a)\cos(\delta_b-\mu_b)  +(1-\nu) p_C \cos(\frac{\theta}{2})\\
& + \nu \sin(\delta_a)\sin(\delta_b) +\nu \sin(\mu_a)\sin(\mu_b)\\
& + (1-\nu) (1-p_C)(\frac{\cos(\theta)}{2} \left( \cos(\delta_b - \mu_b) + \cos(\delta_b + \mu_b) \right) +\sin(\theta) \sin(\delta_b)) .
\end{split}
\ee\\

Next we upper-bound the sines in the expression above with 
\be
\sin(x)\leq x \qquad \text{for}\quad x \geq 0.
\ee
While for the cosines we use the inequality
\be
\cos(x)\leq 1- \frac{1-\cos(\Omega)}{\Omega^2} x^2 \quad \text{for}\quad x\in[0,\Omega],
\ee
and the less trivial one (cf. \App{ref:trigbound})
\be
\cos(x)\cos(y) \leq 1- \frac{\sin^2(\Omega)}{2\Omega^2}(x^2 +y^2) \qquad \text{for} \quad (x,y)\in [0,\Omega]\times[0,\Omega],
\ee
valid for $\Omega\leq \pi$. This last expression allows us to bound the terms
\be
\begin{split}
\cos (\delta_a+\mu_a)\cos(\delta_b+\mu_b) & \leq 1- C_1 \big( (\delta_a+\mu_a)^2 + (\delta_b+\mu_b)^2\big)\\
\cos (\delta_a-\mu_a)\cos(\delta_b-\mu_b) & \leq 1- C_2 \big( (\delta_a-\mu_a)^2 + (\delta_b-\mu_b)^2\big)\\
\cos(\delta_a)\cos(\mu_b) & \leq 1 - C_2 (\delta_b^2+\mu_b^2) \\
\frac{\cos(\theta)}{2} \left( \cos(\delta_b - \mu_b) + \cos(\delta_b + \mu_b) \right) & \leq 1- \frac{1}{2}(C_2 (\theta^2+(\delta_b-\mu_b)^2) + C_1  (\theta^2+(\delta_b+\mu_b)^2))
\end{split}
\ee
with $C_1=\frac{\sin^2(\frac{3 \pi}{16})}{2 (\frac{3 \pi}{16})^2}$ and $C_2=\frac{\sin^2(\frac{ \pi}{8})}{2 (\frac{ \pi}{8})^2}$. Plugging all the inequalities in our objective function gives
\be
\begin{split}
\epsilon_\rho^+ \leq & \nu(1 - \frac{C_1}{2} ((\delta_a + \mu_a)^2 +  (\delta_b + \mu_b)^2) - \frac{C_2}{2} ((\delta_a - \mu_a)^2 +  (\delta_b - \mu_b)^2) + \delta_a\delta_b + \mu_a\mu_b) \\
& +  (1-\nu)(p_C ( 1 - C_3\left(\frac{\theta}{2}\right)^2) +  (1-p_C)(1-\frac{1}{2}(C_2 (\theta^2+(\delta_b-\mu_b)^2) + C_1  (\theta^2+(\delta_b+\mu_b)^2)) + \theta \delta_b) \\
& = 1 + {\bf r}^T T {\bf r}
\end{split}
\ee
with $C_3 = \frac{1-\cos(\frac{\pi}{32})}{\left(\frac{\pi}{32}\right)^2}$, ${\bf r} = (\mu_a \, \mu_b \, \delta_a \, \delta_b, \theta)$ and
\begin{equation}
\nonumber
T =
\begin{footnotesize} \left(
\begin{array}{ccccc}
 -\nu  (C_1+C_2) & \nu  & \nu  (C_2-C_1) & 0 & 0 \\
 \nu  & -(C_1 + C_2) (1-(1-\nu) p_C) & 0 & (C_2-C_1) (1-(1-\nu) p_C) & 0 \\
 \nu  (C_2-C_1) & 0 & -\nu  (C_1+C_2) & \nu  & 0 \\
 0 & (C_2-C_1) (1- (1-\nu) p_C) & \nu  & -(C_1 + C_2) (1-(1-\nu) p_C) & (1-\nu ) (1-p_C) \\
 0 & 0 & 0 & (1- \nu) (1-p_C) & \substack{\frac{1}{2} (1-\nu) ((C_1+C_2) (2 p_C -2)\\ - C_3 p_C)}
\end{array}
\right).
\end{footnotesize}
\end{equation}

For the value of $\nu$ and $p_C$ defined in \eq{eq:param}, the maximum eigenvalue of the matrix T above is given by $T \preceq -0.0146097$. Thus, we obtain,
\be
    \epsilon_\rho^+ \leq  1 - 0.0146097 {\bf r}^2 \leq 1
\ee

With the result of the previous section, this shows that the extracted fidelity of the state $\rho_{AB}$ with the singlet does not exceed $1/2$ for all possible local extraction maps.
\end{widetext}

\subsection{Trigonometric bound}
\label{ref:trigbound}

First, we show the following inequality valid for $\Omega \leq 2\pi$
\be
\cos(x)\leq 1- \frac{1-\cos(\Omega)}{\Omega^2} x^2 \quad \text{for}\quad x\in[0,\Omega].
\ee
This inequality can be rewritten as
\be\nonumber
\frac{1-\cos(x)}{x^2}\geq \frac{1-\cos(\Omega)}{\Omega^2} \quad \text{for}\quad 0\leq x \leq \Omega.
\ee
This inequality holds as the function $\frac{1-\cos(x)}{x^2}$ is decreasing in the range $x\in[0,2\pi]$.\\

Second, let us demonstrate the following inequality
\be
\begin{split}
(1-a\, &x^2)(1- b\, y^2)\\
&\leq 1- a(1-\frac{1}{2} b Y^2)x^2 - b(1-\frac{1}{2} a X^2)y^2 \\ &\text{for}\quad (x,y)\in [0,X]\times[0,Y].
\end{split}
\ee
To show this note
\be\nonumber
(1-a\, x^2)(1- b\, y^2) = 1-a\, x^2 -b\, y^2 + ab \, x^2 y^2,
\ee
with
\begin{align}
\nonumber
ab\, x^2 y^2  \\
= & \frac{ab}{4 X^2 Y^2} \big( (Y^2 x^2 +X^2 y^2)^2- (Y^2 x^2 -X^2 y^2)^2\big) \\
\leq & \frac{ab}{4 X^2 Y^2} (Y^2 x^2 +X^2 y^2)^2\\
\leq & \frac{ab}{4 X^2 Y^2} (Y^2 X^2 +X^2 Y^2)(Y^2 x^2 +X^2 y^2) \\
= & \frac{ab}{2}(Y^2 x^2 +X^2 y^2).
\end{align}
Hence
\be
\begin{split}
(&1-a\, x^2)(1- b\, y^2)\\
&\leq 1- a(1-\frac{1}{2} b Y^2)x^2 - b(1-\frac{1}{2} a X^2)y^2,
\end{split}
\ee

In particular, the inequality implies
\be\begin{split}
\cos(x)\cos(y) \leq 1- \frac{\sin^2(\Omega)}{2\Omega^2}(x^2 +y^2) \\ \text{for} \quad (x,y)\in [0,\Omega]\times[0,\Omega].
\end{split}
\ee
\clearpage

\end{document}